\documentclass[10pt]{article}
\usepackage{graphicx,amsfonts}
\newtheorem{thm}{T{\sc HEOREM}}[section]
\newtheorem{lemma}[thm]{L{\sc EMMA}}

\def\bp{\noindent{\it Proof. }}
\def\ep{\noindent{\hfill \fbox{}}}
\def\remark{\noindent{\bf Remark. }}

\def\pic{\rm Pic }
\def\rank{\rm rank }
\newcommand{\ol}[1]{\overline{#1}}
\newcommand{\mapright}[1]{%
   \smash{\mathop{%
   \hbox to 1cm{\rightarrowfill}}\limits^{#1}}}
\newcommand{\mapleft}[1]{%
   \smash{\mathop{%
   \hbox to 1cm{\leftarrowfill}}\limits^{#1}}}
\newcommand{\maplleft}[2]{%
   \smash{\mathop{%
   \hbox to 1cm{\leftarrowfill}}\limits_{#1}^{#2}}}

\begin{document}

\title{Algebraic entropy and the space of initial values for discrete 
dynamical systems}
\author{Tomoyuki Takenawa}

\date{}
\maketitle
\begin{center}
{Graduate School of Mathematical Sciences, University of
Tokyo, Komaba 3-8-1, Meguro-ku, Tokyo 153-8914, Japan}\\
\end{center}

\begin{abstract}
A method to calculate the algebraic entropy of a mapping which can be 
lifted to an isomorphism of a suitable rational surfaces (the space of initial 
values)  are presented. It is shown that the degree of the $n$th iterate of such 
a mapping is given by its action on the Picard group of the space of initial values.
It is also shown that the degree of the $n$th  iterate of every Painlev\'e
equation in sakai's list is at most $O(n^2)$ and therefore its algebraic
entropy is zero.  
\end{abstract}

%%%%%%%%%%%%%%%%%%%%%%
\section{Introduction}

The notion of algebraic entropy was introduced by Hietarinta and Viallet
\cite{hv} in order to test the degree of complexity of successive iterations
of a rational mapping.
The algebraic entropy is defined as $s:=\lim_{n\to \infty}\log(d_n)/n$
where $d_n$ is the degree of the $n$th iterate.
This notion is linked to Arnold's complexity since
the degree of a mapping gives the intersection number of the image
of a line and a hyperplane. While the degree grows exponentially for
a generic mapping, it was shown that it only grows
polynomially  for a large class of integrable mappings
\cite{hv,arnord,bv}. Especially the case of some discrete Painlev\`e equations
are studied by Ohta et. al. \cite{otgr}. 

Let $\varphi_i$ be a birational mapping of ${\mathbb P}^2$ or
${\mathbb P}^1 \times {\mathbb P}^1$. 
A sequence of rational surfaces $X_{i}$ is (or $X_i$ themselves are) called 
the space 
of initial values for the sequence of $\varphi_i$ if each $\varphi_i$ is 
lifted to 
an isomorphism, i.e. bi-holomorphic mapping, from $X_i$ to $X_{i+1}$
\cite{okamoto, takenawa, takenawa2}.  Here, the mapping $\varphi'$ is called a mapping lifted 
from the mapping $\varphi$ if $\varphi'$ coincides with $\varphi$
on any point where $\varphi$ is defined. 
Such a mapping induces an action on the Picard group of its space
of initial values. Here, the Picard group of a rational surface $X$ is the 
group 
of isomorphism classes of invertible sheaves on $X$ and
it is isomorphic to the group of linear equivalent classes of divisors on $X$.

In this paper we present a method to calculate the degree of the
$n$th iterate of the sequence of mappings which has the space of initial values.
We also show an example of calculation and simplify the method
by considering invariant sublattices. We apply our method to the discrete 
Painlve\'e equations in Sakai's list \cite{sakai} and show that for all 
of them the degrees grow at most in the order $n^2$.

The discrete Painlev\'{e} equations were found by many authors \cite{rgh,js}
and have been extensively studied. Recently it was shown by 
Sakai \cite{sakai} that all
of (from the point of view of symmetries) these are
obtained by studying rational surfaces in connection with the extended
affine Weyl groups. 

Surfaces obtained by successive blow-ups \cite{hartshorne} of ${\mathbb P}^2$ or
${\mathbb P}^1 \times {\mathbb P}^1$ have been studied by several authors 
by means of connections between the Weyl groups and the groups of 
Cremona isometries on the 
Picard group of the surfaces \cite{cd,dolgachev,dolgachev2}.
Here, a Cremona isometry is an isomorphism of the Picard group such that
a) it preserves the intersection number of any pair of divisors, 
b) it preserves the canonical divisor $K_X$ and 
c) it leaves the set of effective classes of divisors invariant.
In the case where $9$ points (in the case of ${\mathbb P}^2$, $8$ points
in the case of ${\mathbb P}^1 \times {\mathbb P}^1$) are blown up, 
if the points are in general position the group of Cremona 
isometries becomes isomorphic with 
an extension of the Weyl group of type $E_8^{(1)}$. 
In case the $9$ points are not in general position,
the classification of  connections between the group of Cremona isometries
and the extended affine Weyl groups
was first studied by Looijenga \cite{looijenga} and
more generally by Sakai.
Birational (bi-meromorphic) mappings on ${\mathbb P}^2$ (or
${\mathbb P}^1 \times {\mathbb P}^1$) are obtained by interchanging 
the procedure of blow downs.  Discrete
Painlev\'{e} equations are recovered as the birational mappings
corresponding to the translations of the affine Weyl groups.

In Section~2, we show a method to calculate the degree of the $n$th iterate.
Considering the intersection numbers of
divisors it is shown that the degree is given by the $n$th power of 
a matrix given by the action of the mapping on the Picard group.   

In Section~3, we present an example of calculation. We apply our method
to the mapping which was found by Hietarinta and Viallet \cite{hv}
and whose space of initial values are obtained by $14$ blow-ups from
${\mathbb P}^1 \times {\mathbb P}^1$ \cite{takenawa}. 
We simplify calculation using the root systems
associated with the symmetries of their space of initial values. 

In Section~4, applying our method for discrete Painlev\'{e} equations, 
we show the 
degrees of $n$th iterate are at most $O(n^2)$.

%%%%%%%%%%%%%%%%

\section{Algebraic entropy and intersection numbers}

We define the degree of polynomial on ${\mathbb P}^1(=
{\mathbb P}^1({\mathbb C})):$
$f(t)=\sum_{m}a_{t}t^m$ as
$$\deg_t(f(t))=\max\{m~;~a_m\neq 0\},$$ 
where $t$ is a coordinate of ${\mathbb C} \subset {\mathbb P}^1= 
{\mathbb C}\cup \{\infty\}$.
Similarly we define the degree of a polynomial on ${\mathbb P}^1\times {\mathbb P}^1:$
$f(x,y)=\sum_{m,n}a_{m,n}x^my^n$
as $$\deg(f(x,y))=\max\{m+n~;~a_{m,n}\neq 0\}.$$ 
The degree of an irreducible rational function 
$P(x,y)=f(x,y)/g(x,y)$, where 
$f(x,y)$ and $g(x,y)$ are polynomials, 
is defined by
$$\deg(P)=\max\{\deg f(x,y),\deg g(x,y)\}.$$
The degree of a mapping $\varphi:{\mathbb P}^1\times {\mathbb P}^1
\to {\mathbb P}^1\times {\mathbb P}^1$, $(x,y)\mapsto(P(x,y),Q(x,y))$, where
$P(x,y)$ and $Q(x,y)$ are rational functions, is defined by
$$\deg(\varphi)=\max\{\deg P(x,y),\deg Q(x,y)\}$$
and similarly $\deg_t(\varphi)$ is defined by the degree about $t$.

The algebraic entropy $h(\varphi)$, where 
$\varphi$ is a mapping from  ${\mathbb P}^1\times {\mathbb P}^1$ to itself, 
is defined by 
$$h(\varphi)=\lim_{n\to\infty}\frac{1}{n}\log\deg(\varphi^n)$$
if the limit exists. \\

\remark
If one would prefer to discuss the mapping in ${\mathbb P}^2$ instead of
${\mathbb P}^1\times {\mathbb P}^1$, it is sufficient to note that
we can relate a mapping 
$\varphi'$: $(X,Y,Z)\in {\mathbb P}^2 \mapsto 
(\ol{X},\ol{Y},\ol{Z}) \in{\mathbb P}^2$ 
with a mapping 
$\varphi$: $(x,y)\in{\mathbb P}^1\times {\mathbb P}^1 \mapsto 
(\ol{x},\ol{y})\in {\mathbb P}^1\times {\mathbb P}^1$ 
by using the relations $x=X/Z,y=Y/Z$ and $\ol{x}=\ol{X}/\ol{Z},
\ol{y}=\ol{Y}/\ol{Z}$ 
and by reducing to a common denominator.
We denote the $n$th iterate of $\varphi'$ by
$(f_n(X,Y,Z),g_n(X,Y,Z),h_n(X,Y,Z))$ where $f_n,g_n,h_n$ are polynomials
with the same degree and should be simplified if possible.
The algebraic entropy $h(\varphi)$ then coincides with $h(\varphi')$,
where $h(\varphi')$ is defined by 
$\deg(\varphi'^n)=\deg f_n(=\deg g_n=\deg h_n)$ 
and $\lim_{n\to\infty}\deg(\varphi')$.\\

We show that we can calculate the degree of  
the $n$th iterate and thus the entropy of mapping
by using the theory of intersection numbers.

Let $\{X_i\}$ be a sequence of rational surfaces obtained by successive $m$ blow-ups 
from ${\mathbb P}^1\times {\mathbb P}^1$ and 
let $\varphi_i(x,y)$ be an isomorphism from $X_{i-1}$ to $X_i$.
We write the action of 
$$\varphi_n\circ\varphi_{n-1}\circ\cdots\circ\varphi_1$$ 
on ${\mathbb P}^1\times {\mathbb P}^1$
as $(P_n(x,y),Q_n(x,y))$.

We denote the linear equivalent classes of total transform of
$x={\rm constant},$ (or $y={\rm constant}$) on $X$ by $H_0$ (or $H_1$ respectively)
and the linear equivalent classes of total transform of the point 
of the $i$-th blow-up by $E_i$. 
From \cite{hartshorne} we know that the Picard group of $X$, Pic($X$),
is
$${\rm Pic}(X)= {\mathbb Z}H_0 + {\mathbb Z}H_1+  {\mathbb Z}E_1+\cdots+{\mathbb 
Z}E_{m}$$
and the intersection form, i.e. the intersection numbers
of pairs of base elements, is
\begin{eqnarray}\label{isn}
H_i \cdot H_j = 1-\delta_{i,j},~ E_k\cdot E_l= -\delta_{k,l},~
H_i \cdot E_k=0~
\end{eqnarray}
where $\delta_{i,j}$ is $1$ if $i=j$ and $0$ if $i\neq j,$ and
the intersection numbers of any pairs of divisors are given by
their linear combinations. \\

\remark
Let $X$ be a rational surface.
It is known that Pic($X$), the group of isomorphism classes of 
invertible sheaves of $X$, is isomorphic to the following groups. \\
i)The group of linear equivalent classes of divisors on $X$.\\
ii)The group of numerically equivalent classes of divisors on $X$,
where divisors $D$ and $D'$ on $X$ are numerically equivalent
if and only if for any divisors $D''$ on $X$, $D\cdot D''=D'\cdot D''$
holds. \\
Hence we identify them in this paper. \\

Let us define the curve $L$ in $X_0$ as $y= c_1 x+c_2$ where 
$c_1,c_2 \in {\mathbb C}$ are nonzero constants.
Notice that
\begin{eqnarray}\label{txy} 
\deg_t(P_n(t,c_1t+c_2)) =\deg(P_n(x,y))
\end{eqnarray}
holds for generic $c_1$ and $c_2$.

By the fundamental theorem of algebra, 
$\deg_t(P_n(t,c_1t+c_2))$  
coincides with the intersection number of the curve
$(x,y)=(P_n(t,c_1t+c_2),Q_n(t,c_1t+c_2)$ and the curve 
$x=d$ in ${\mathbb P}^1 \times {\mathbb P}^1$,
where $d\in {\mathbb C}$ is a constant. 
The class of the curve $x=d$ is expressed as $H_0$ in $\pic(X)$
and hence from (\ref{isn}) this intersection number
coincides with the coefficient of $H_1$ of the class
of the curve $\varphi^n(L)$ for nonzero constants $c_1$ and $c_2$.
Analogously, the intersection number of the curve
$(x,y)=(P_n(t,c_1t+c_2),Q_n(t,c_1t+c_2)$ and the curve $y=d$  
coincides with the coefficient of $H_0$ of the class
of the curve $\varphi^n(L)$ for any nonzero constants $c_1$ and $c_2$.

Notice that
for any isomorphism $\theta$ 
from the rational surface $X$
to the rational surface $X'$ and any divisor $D$,  the relation
$$[\theta(D)]=\theta([D])$$
holds, where $[*]$ means the class of $*$ 
and in the right hand side
$\theta$ is the linear operator on Pic($X$)
($=$ Pic($X'$)) which is induced by the isomorphism $\theta$.

In our case, we have
\begin{eqnarray*}
[\varphi_n\circ\cdots\circ\varphi_1(L)]= 
\varphi_n\circ\cdots\circ\varphi_1([L])
\end{eqnarray*}
and therefore writing the coefficients of $H_0$ and $H_1$ of 
$\varphi_n\circ\cdots\circ\varphi_1([L])$ 
as $h_n^0,h_n^1$, 
we have the relation 
\begin{eqnarray*}
\deg_t(P_n(t,c_1t+c_2))=h_n^1&
\deg_t(Q_n(t,c_1t+c_2))=h_n^0
\end{eqnarray*}
for any nonzero constants $c_1,c_2\in {\mathbb C}$.
On the other hand it can be seen that
the relation (\ref{txy}) holds for $(c_1,c_2)=(c_1^0,c_2^0)$ such that
$[L]$ is invariant under infinitesimal change of $(c_1,c_2)$ around 
$(c_1^0,c_2^0)$,
i.e. $[L]$ is generic at the point $(c_1,c_2)=(c_1^0,c_2^0)$ 
for the parameter $(c_1,c_2)$.
(Suppose that $P_n(t,c_1t+c_2)$ would suddenly be simplified 
at this point and that $[L]$ is invariant under infinitesimal 
change of $(c_1,c_2)$ around $(c_1^0,c_2^0)$.
The intersection number of the curve $x=P_n(t,c_1t+c_2)$ and 
the curve $x=d$  
would then change but $\varphi([L])$ itself would still be invariant, 
which leads to a contradiction. Similarly for the case of 
$Q_n(t,c_1t+c_2)$.)

Consequently we have the following theorem.
\begin{thm}
Let $\{X_i\}$ be a sequence of rational surfaces obtained by blow 
ups from ${\mathbb P}^1\times {\mathbb P}^1$ and 
let $\varphi_i(x,y)$ be an isomorphism from $X_{i-1}$ to $X_i$.
We denote the action of 
$\varphi_n\circ\cdots\circ\varphi_1$ 
on ${\mathbb P}^1\times {\mathbb P}^1$
by $(P_n(x,y),Q_n(x,y))$.
Let $[L]$ be the class of curve $y=c_1x+c_2$ in $X_0$ such that 
$[L]$ is generic 
and let $h_n^0,h_n^1$ be the coefficients of $H_0$ and $H_1$ of  
$\varphi_n\circ\cdots\circ\varphi_1([L])$. 
The formula
\begin{eqnarray*}
\deg(P_n(x,y))=h_n^1&
\deg(Q_n(x,y))=h_n^0.
\end{eqnarray*}
then holds.
\end{thm}

\remark
As before if $\{X_i'\}$ is a sequence of rational surfaces obtained by blow 
ups from ${\mathbb P}^2$ (instead of ${\mathbb P}^1\times{\mathbb P}^1$) and 
$\varphi_i(x,y)$ is an isomorphism from $X_{i-1}'$ to $X_i'$,
we can consider the degree of the mapping 
$$\varphi_n'\circ\cdots\circ\varphi_1' 
: {\mathbb P}^2 \to {\mathbb P}^2.$$
We denote the class of a curve $aX+bY+cZ=0$ in ${\mathbb P}^2$ 
by ${\cal E}$. Notice that
${\cal E}$ is always generic for parameters $a,b,c$ in the rational surface $X_0$.
The intersection form is
${\cal E}\cdot{\cal E}=1,~~ E_i\cdot E_j=-\delta_{i,j},
~~{\cal E}\cdot E_i=0$.
Similar to the case of ${\mathbb P}^1 \times {\mathbb P}^1$,
we have the fact that the degree of $\varphi_n'\circ\cdots\circ\varphi_1'$
coincides with 
the coefficient of ${\cal E}$ of 
$\varphi_n'\circ\cdots\circ\varphi_1'({\cal E})$.\\

\section{An example and simplification}

We consider the following equation found by Hietarinta and Viallet \cite{hv} (we 
denote it as the HV eq. in this paper):
\begin{eqnarray}\label{hv}
\varphi:&{\mathbb P}^1 \times {\mathbb P}^1 & \to ~ {\mathbb P}^1 \times
{\mathbb P}^1  \nonumber \\
&\left(\begin{array}{c}
x_n\\y_n \end{array}\right)
& \mapsto ~
\left(\begin{array}{c}
x_{n+1}\\y_{n+1} \end{array}\right)
= \left(\begin{array}{c}
y_n\\ -x_n+y_n+a/y_n^2 \end{array}\right) \label{hv} 
\end{eqnarray}
where $a \in {\mathbb C}$ is a nonzero constant.
It is known that the algebraic entropy of the HV eq. $\varphi$ 
is equal to $\log (3+\sqrt{5})/2$.
Here we shall recover the algebraic entropy of the HV eq. 
by using the theory of intersection numbers.

The HV eq. can be lifted to an automorphism of a rational surfaces $X$ obtained 
by successive $14$ blow-ups
from ${\mathbb P}^1 \times {\mathbb P}^1$ \cite{takenawa}.
Hence its Picard group is 
$${\rm Pic}(X)= {\mathbb Z}H_0 + {\mathbb Z}H_1+  {\mathbb Z}E_1+\cdots+{\mathbb 
Z}E_{14},$$
where total transforms of the points of blow-ups as follows:
{\footnotesize
\begin{eqnarray*}
E_1: (1/x,y)=(0,0) && E_2: \left(\frac{1}{xy},y\right)=(0,0)\\
E_3: \left(\frac{1}{xy},xy^2\right)=(0,a)&&
E_4: \left(\frac{1}{xy}, xy(xy^2-a)\right)=(0,0)\\
E_5: (x,1/y) =(0,0)&& E_6: \left(x,\frac{1}{xy}\right)=(0,0)\\
E_7: \left(x^2 y,\frac{1}{xy}\right) =(a,0) &&
E_8: \left(xy(x^2 y-a),\frac{1}{xy}\right)=(0,0)\\
E_9:(1/x,1/y)=(0,0)&& E_{10}: \left(\frac{1}{x},\frac{x}{y}\right)
=(0,1)\\
E_{11}:  \left(\frac{1}{x},x(\frac{x}{y}-1)\right)=(0,0)
&& E_{12}: \left(\frac{1}{x^2(x/y-1)},  x(\frac{x}{y}-1)\right)=(0,0)\\
E_{13}: \left(\frac{1}{x^2(x/y-1)},  x^3(\frac{x}{y}-1)^2\right)=(0,a) &&
E_{14}: \left(\frac{1}{x^2(x/y-1)},
x^2(\frac{x}{y}-1)(x^3(\frac{x}{y}-1)^2-a)\right)=(0,0).
\end{eqnarray*}
 }
Its action on the Picard group is  
\begin{eqnarray}\label{actp}
\left(
\begin{array}{c}
H_0\\
H_1,~~E_1,~~ E_2\\
E_3,~~E_4,~~E_5,~~E_6\\
E_7,~~E_8,~~E_9,~~E_{10}\\
E_{11},~~E_{12},~~E_{13},~~E_{14}
\end{array}\right)
\to
\left( \begin{array}{c}
3H_0+H_1-E_5-E_6-E_7-E_8-E_9-E_{10}\\
H_0,~~H_0-E_8,~~H_0-E_7 \\
H_0-E_6,~~H_0-E_5,~~E_{11},~~E_{12}\\
E_{13},~~E_{14},~~H_0-E_{10},~~H_0-E_9\\
E_1,~~ E_2,~~E_3,~~E_4
\end{array}\right)
\end{eqnarray}
(this table means $\overline{H_0}=3H_0+H_1-E_5-E_6-E_7-E_8-E_9-E_{10},$
$\overline{H_1}=H_0,$  $\overline{E_1}=H_0-E_8$ and so on)
and their linear combinations. \\ 

Notice that 
(\ref{actp}) means a change of bases.
Actually by fixing the basis of Pic($X$)
as $\{ H_0,H_1,E_1,E_2,\cdots,E_{14}\}$,
this table can be expressed
by the following matrix as the action from the left hand side on the space of 
coefficients of basis.
{\footnotesize
\begin{eqnarray}\label{actpm}
\left(
\begin{array}{cccccccccccccccc}
3&1&1&1&1&1&0&0&0&0&1&1&0&0&0&0\\
1&0&0&0&0&0&0&0&0&0&0&0&0&0&0&0\\
0&0&0&0&0&0&0&0&0&0&0&0&1&0&0&0\\
0&0&0&0&0&0&0&0&0&0&0&0&0&1&0&0\\
0&0&0&0&0&0&0&0&0&0&0&0&0&0&1&0\\
0&0&0&0&0&0&0&0&0&0&0&0&0&0&0&1\\
-1&0&0&0&0&-1&0&0&0&0&0&0&0&0&0&0\\
-1&0&0&0&-1&0&0&0&0&0&0&0&0&0&0&0\\
-1&0&0&-1&0&0&0&0&0&0&0&0&0&0&0&0\\
-1&0&-1&0&0&0&0&0&0&0&0&0&0&0&0&0\\
-1&0&0&0&0&0&0&0&0&0&0&-1&0&0&0&0\\
-1&0&0&0&0&0&0&0&0&0&-1&0&0&0&0&0\\
0&0&0&0&0&0&1&0&0&0&0&0&0&0&0&0\\
0&0&0&0&0&0&0&1&0&0&0&0&0&0&0&0\\
0&0&0&0&0&0&0&0&1&0&0&0&0&0&0&0\\
0&0&0&0&0&0&0&0&0&1&0&0&0&0&0&0
\end{array}\right)
\end{eqnarray}}

The curve $L:x=c_1y+c_2$, where $c_1,c_2\in {\mathbb C}$ are nonzero constants,is 
expressed by $H_0+H_1-E_9$ in Pic($X$) if $c_1\neq 1$.
This fact is easily calculated from the fact that $L$ has 
intersections only with $H_0,H_1$ and $E_9$ at one time.

The action of $\varphi$ on Pic($X$) is given by (\ref{actp}) or (\ref{actpm}). 
Hence the algebraic entropy of the HV eq.,
$\lim_{n\to\infty}\frac{1}{n} \log\max\{h_n^0,h_n^1\}$,
can be shown to be equal to (by diagonalization of the matrix (\ref{actpm})):
\begin{eqnarray*}
\log\max\{ |\mbox{ eigenvalues of } (\ref{actpm})| \}&=
&\log\frac{3+\sqrt{5}}{2}.
\end{eqnarray*}

On the level of the mapping itself, the degrees can be calculated as follows:
\begin{eqnarray}
(x,y)&\mapright{\varphi}&(y,\frac{-xy^2+y^3+a}{y^2})
~\mapright{\varphi}~(\deg 3, \deg9)\nonumber \\
&\mapright{\varphi}&(\deg9,\deg25)
~\mapright{\varphi}~(\deg 25, \deg67)~\mapright{\varphi} \cdots.~ \label{rei1}
\end{eqnarray}
On the other hand, the intersection numbers can be calculated 
by (\ref{actp}) or (\ref{actpm}) as follows:
\begin{eqnarray*}
H_0+H_1-E_9&\mapright{\varphi}&3H_0+H_1-E_5-E_6-E_7-E_8-E_9\\
&\mapright{\varphi}& 9H_0+3H_1+\cdots+(-3)E_9+\cdots\\
&\mapright{\varphi}& 25H_0+9H_1+\cdots+(-7)E_9+\cdots\\
&\mapright{\varphi}& 67H_0+25H_1+\cdots+(-19)E_9+\cdots\\
&\mapright{\varphi}& \cdots 
\end{eqnarray*}
which actually coincides with (\ref{rei1}).

By the corresponding mapping in ${\mathbb P}^2$,
the degrees are
\begin{eqnarray}\nonumber
(X,Y,Z)&\mapright{\varphi'}&(Y^3,-X Y^2+Y^3+a Z^3,Y^2Z)
~\mapright{\varphi'}~\deg 9 \\
&\mapright{\varphi'}&
\deg27~\mapright{\varphi'}~\deg73 ~\mapright{\varphi'}\cdots. \label{rei2}
\end{eqnarray}
Using the correspondence ${\cal E}=H_0+H_1-E_9$ (it is 
shown that ${\cal E}:aX+bY+cZ=0$ actually has this correspondence
in appendix A), we have that the curve has the property
\begin{eqnarray*}
\varphi^n({\cal E})\cdot{\cal E}
&=& \varphi^n(H_0+H_1-E_9)\cdot(H_0+H_1-E_9)\\ 
&=&(h_0H_0+h_1H_1+e_1E_1+\cdots+e_{14}E_{14})\cdot(H_0+H_1-E_9)\\
&=& h_0+h_1+e_9,
\end{eqnarray*}
where $e_i\in {\mathbb Z}$ is the coefficient of $E_i$.
Hence we have the sequence of the coefficients of ${\cal E}$ as
\begin{eqnarray*}
1~\mapright{\varphi'}~3~\mapright{\varphi'}~9~\mapright{\varphi'}~27~
\mapright{\varphi'}~73~\mapright{\varphi'}\cdots,
\end{eqnarray*}
which coincides with (\ref{rei2}).

\medskip

Next we consider simplification of our method.
The anti-canonical divisor $-K_X$ can be reduced uniquely \cite{takenawa2} 
to prime divisors as
\begin{eqnarray}\label{dkx}
D_0+2D_1+D_2+D_3+ 2D_4+D_5+3D_6+D_7+2D_8+D_9+2D_{10}+2D_{11}+2D_{12}
\end{eqnarray}
where 
$$D_0= E_1-E_2 ~~~D_1= E_2-E_3 ~~~D_2=E_3-E_4 ~~~D_3=E_5-E_6 ~~~
 D_4=E_6-E_7$$
$$D_5=E_7-E_8 ~~~D_6=E_9-E_{10} ~~~D_7=E_{11}-E_{12} ~~~
D_8=E_{12}-E_{13} ~~~D_9=E_{13}-E_{14}$$
$$D_{10}=H_0-E_1-E_2-E_9 ~~~ D_{11}=H_1-E_5-E_6-E_9 ~~~
 D_{12}=E_{10}-E_{11}-E_{12}.$$
We denote sub-lattice of the Picard group $\sum_{i=0}^{12}{\mathbb Z}D_i$ as $<D_i>$.
Let $<\alpha_i>$ be orthogonal complement of $<D_i>$ and
let $\{\alpha_1,\alpha_2,\alpha_3\}$ be its basis. Notice that
$\varphi$ preserves $<D_i>$ and $<\alpha_i>$ because 
its action on the Picard group is a Cremona isometry.

The matrix (\ref{actpm}) is an expression of the action $\varphi$ on Pic($X$) 
by the basis\\
$\{H_0, H_1, E_1, \cdots, E_{14}\}$. But  
$\{D_0,D_1,\cdots,D_{12},\alpha_1,\alpha_2,\alpha_3\}$ is a
better basis for calculation of the degree of 
$\varphi_n$ (we may consider $\pic(X)$ to be a vector space on
${\mathbb C}$ instead of ${\mathbb Z}$ module for this purpose).  
The reason is that $<D_i>$ and $<\alpha_i>$ are 
eigenspaces of $\varphi$ and compliment each other and
moreover the action of $\varphi$ on $<D_i>$ is just
a permutation. Hence it is enough to investigate the action on
$<\alpha_i>$ in order to know the level of growth of $\deg(\varphi_n)$.  

Actually, by taking a basis of $<\alpha_i>$ as 
\begin{eqnarray*}
\alpha_1&=&2H_1-E_1-E_2-E_3-E_4,\nonumber\\
\alpha_2&=&2H_0-E_5-E_6-E_7-E_8,\\
\alpha_3&=&2H_0+2H_1-2E_9-2E_{10}-E_{11}-E_{12}-E_{13}-E_{14}\nonumber
\end{eqnarray*}
(the fact is, this basis is the basis of root system by regarding the 
$<\alpha_i>$ and intersection form as 
the root lattice and the bilinear form respectively)
and writing an element of $<\alpha_i>$ as 
$r_1\alpha_1+r_2\alpha_2+r_3\alpha_3$,  
the action of 
$\varphi$ on
$<\alpha_i>$ is expressed as 
\begin{eqnarray}
\left( \begin{array}{c}
\ol{r_1} \\ \ol{r_2} \\ \ol{r_2}
\end{array} \right)
&=&
\left( \begin{array}{ccc}
0&0&1\\
-1&2&2\\
0&1&0\end{array} \right)
\left( \begin{array}{c}
r_1 \\ r_2 \\ r_2
\end{array} \right)
\end{eqnarray}
and the absolute values of its eigenvalues are $1,(3\pm\sqrt{5})/2$.\\

\remark 
The non-autonomous version is as follows:
\begin{eqnarray}
\varphi
&&:(x,y;a_1,a_2,a_3,a_4,a_5,a_6,a_7)\nonumber\\
&\mapsto&(y,-x+y+a_5+\frac{a_1}{y^2}+\frac{a_2}{a_1y}~;~\\
&& a_6,-a_7+2a_5^2a_6+\frac{2a_2a_6}{a_1},a_1,a_2,-a_5,a_3,
a_4+2a_3a_5^2-\frac{2a_2a_3}{a_1}), \nonumber
\end{eqnarray}
where
$a_i\in {\mathbb C}$ and $a_1,a_3,a_6$ are nonzero and an over-line means
the value of the image by the mapping\cite{takenawa}.
In this case the coefficients of $H_i$ and $E_i$ do not change
and therefore the degrees and the algebraic entropy do not change, 
since its action on the Picard group is identical with the  
action of the original autonomous version.

\section{The growth of degree of discrete Painlev\'{e} equations}

It is shown by Sakai \cite{sakai} that the discrete Painlev\'{e} equations 
can be obtained by the following method.

Let $X$ be a rational surface obtained by blow-ups from ${\mathbb P}^2$
such that its anti-canonical divisor 
$-K_X$ ($=3{\cal E}-E_1-\cdots-E_9$) is uniquely decomposed
in prime divisors as  $-K_X=\sum_{i=1}^I m_iD_i$ and satisfies 
$K_X\cdot D_i=0$
for all $i$. This implies that $K_X\cdot K_X=0$ and therefore $X$
is obtained by $9$ points blow-ups from ${\mathbb P}^2$ and hence 
$ \rank {\rm Pic}(X)=10$.
One can classify
such surfaces according to the type (denoted by $R$) 
of Dynkin diagram formed by
the $D_i$ (the lattice of $R$ is a sub-lattice of the lattice of $E_8^{(1)}$).    

The Cremona isometries of $X$ preserve the sub-lattice $<D_i>$
and its orthogonal sub-lattice with respect to the intersection form.
By taking a suitable basis of the orthogonal lattice,
$\{\alpha_1,\alpha_2,\cdots,\alpha_J\}$,
and by regarding $<\alpha_j>$ and the intersection form as the root lattice
and the bilinear form respectively,
it becomes the basis of an extended affine Weyl group
and moreover $\alpha_j\cdot\alpha_j$ does not depend on $j$.
All the actions of these extended affine Weyl groups on  $<\alpha_j>$
are uniquely extended to the actions on $\pic(X)$ as Cremona isometries.
Notice that the intersection number of $\alpha_j$ and
$K_X$ is zero, since  $\alpha_j\cdot K_X$ $=\alpha_j \cdot \sum m_iD_i=0$.
Similar to the case of the HV eq., every discrete Painlev\'e equation acts
on $\{D_i\}$ just as a permutation (this fact follows from the uniqueness of
decomposition of the anti-canonical divisor and the definition of Cremona
isometry).

The group of Cremona isometries of $X$ is isomorphic to 
the extended affine Weyl group
and each element can be realized as a Cremona transformation,
i.e. birational mapping, 
on ${\mathbb P}^2$.
Each of the discrete Painlev\'{e} equations corresponds to a 
translation of extended affine Weyl group.

The Cartan matrixes of these affine Weyl groups are symmetric
and $-K_X$  becomes the canonical 
central element (and
also becomes $\delta$, see \S~6.2 \S~6.4 in \cite{kac}).
Hence the action of Painlev\'{e} equation on the orthogonal lattice 
$<\alpha_j>$ 
is expressed as 
\begin{eqnarray}\label{trans}
(\alpha_1,\alpha_2,\cdots,\alpha_J)\mapsto
(\alpha_1+k_1K_X,\alpha_2 + k_2K_X,\cdots,\alpha_J+k_JK_X)
\end{eqnarray}
where $k_j\in {\mathbb Z}$ and $\sum k_j=0$.

\begin{lemma}\label{rank}
Let $X$, $D_i$ and $\alpha_j$ be as mentioned above. 
The following formula with respect to the rank:  
$$\rank<D_1,\cdots,D_I,\alpha_1,\cdots,\alpha_J>=9$$
holds.
\end{lemma}

\bp Notice that $\{D_1,\cdots,D_I\}$ or $\{\alpha_1,\cdots,\alpha_J\}$
are linearly independent.
Suppose $\sum d_iD_i+ \sum r_j\alpha_j=0$, where $d_i,r_j\in{\mathbb C}$.
We have $F:=-\sum d_iD_i=\sum r_j\alpha_j\in <D_i>\cap<\alpha_j>$.
Since $F$ is an element of $<D_i>$,
$\alpha_i\cdot(\sum r_j \alpha_j)=0$ holds for all $1\leq i \leq J$.
Here the Cartan matrix of the Weyl group is
$C:=(c_{i,j})_{1\leq i,j \leq J}$:
$$c_{i,j}= \frac{2\alpha_i\cdot \alpha_j}{\alpha_i\cdot\alpha_i}$$
and $\alpha_i\cdot\alpha_i$ does not depend on $i$.
Hence it implies 
\begin{eqnarray}\label{aa}
C {\bf r}=0,
\end{eqnarray} 
where ${\bf r}=(r_1,\cdots,r_J)$.
The corank of Cartan matrix of affine type
is $1$. Hence we obtain $F \in {\mathbb Z}K_X$. It implies the fact that 
the corank
of $<D_1,\cdots,D_I,\alpha_1,\cdots,\alpha_J>$ is $1$.
\ep  \\

Let $E_9$ be an exceptional curve, where ``$9$'' means the last blow-up.

\begin{lemma}\label{522}
$$\{D_1,\cdots,D_I,\alpha_1,\cdots,\alpha_J,K_X,E_9\}$$
is a basis of Pic($X$).\end{lemma}
Of course these elements are not 
independent.\\

\bp
Suppose $E_9=\sum d_i D_i + \sum r_j \alpha_j$, where $d_i,r_j\in{\mathbb C}$.
Multiplying this equation by $K_X$, we find
$-1=0$. The claim of lemma follows from Lemma~\ref{rank}.\ep\\

Let $T$ be a discrete Painlev\'{e} equation.
Since $T$ acts on $\{D_i\}$ just as a permutation,
there exists $l$ such that $T^l$ acts on  $\{D_i\}$ as the identity.

\begin{lemma}\label{523}
There exist integers $z_1,z_2,\dots,z_J$ such that
\begin{eqnarray*}
T^l(E_9)=E_9+\sum z_j \alpha_j
\end{eqnarray*}
holds.
\end{lemma}

\bp Notice that $E_9$ has an intersection with only one of $\{D_i\}$ and
without loss of generality we can assume $E_9\cdot D_1=1$.
The system of equations $T^l(E_9)\cdot D_1=1$, 
$T^l(E_9)\cdot D_i=0 ~(i=2,\dots I)$ is linear. Hence the solutions 
of this system are  $T^l(E_9)=E_9+\sum {\mathbb C} \alpha_j$. 
Of course $T^l(E_9)$ must be an element of Pic($X$) and therefore 
the coefficients must be integers. 
\ep\\

By (\ref{trans}), Lemma~\ref{522} and Lemma~\ref{523} the action of $T^l$ on Pic($X$)
is expressed as
\begin{eqnarray*}
&&d_1D_1+\cdots+d_ID_I+r_1\alpha_1+\cdots+r_J\alpha_J+kK_X+eE_9 \\
&\mapsto&
d_1D_1+\cdots+d_ID_I+(r_1+e z_1)\alpha_1+\cdots+(r_J+e z_J)\alpha_J \\
&& +(k+ l r_1 k_1+\dots+l r_J k_J) K_X + e E_9
\end{eqnarray*}
where $d_i,r_j,k,e\in{\mathbb Z}$.
This action is written by the matrix
\begin{eqnarray}
A&:=&
\left[  
\begin{array}{ccc|ccc|c|c}
1&      & &    &       &    & &      \\
 &\ddots& &    &       &    & &      \\
0&      &1&    &       &    & &      \\ \hline
 &      & &   1&       &    & &  z_1\\
 &      & &    &\ddots &    & &\vdots\\
 &      & &    &       &   1& &  z_J\\ \hline
 &      & &lk_1&\cdots &lk_J&1&     \\ \hline
 &      & &    &       &    & &     1
\end{array} \right],
\end{eqnarray}
where a blank means $0$.

The matrix $A^s$, where $s\in {\mathbb N}$ is
\begin{eqnarray}
A^s&:=&
\left[  
\begin{array}{ccc|ccc|c|c}
1&      & &    &       &    & &      \\
 &\ddots& &    &       &    & &      \\
0&      &1&    &       &    & &      \\ \hline
 &      & &   1&       &    & & s z_1\\
 &      & &    &\ddots &    & &\vdots\\
 &      & &    &       &   1& & s z_J\\ \hline
 &      & &s lk_1&\cdots &s lk_J&1& *_s \\ \hline
 &      & &    &       &    & &     1
\end{array} \right],
\end{eqnarray}
where $*_s=\frac{1}{2}s(s-1)\sum l k_j z_j$.

Let us start with ${\cal E}\in {\rm Pic}(X)$ and let
$d_1D_1+\cdots+d_ID_I+r_1\alpha_1+\cdots+r_J\alpha_J+kK_X+eE_9$
be an expression of ${\cal E}$.
We obtain the following theorem.

\begin{thm}
For all discrete Painlev\'{e} equations the order of degree 
of the $n$th iterate is at most $O(n^2)$.
\end{thm}

\bp The degree of the Painlev\'{e} equation $T$ as a birational mapping
of ${\mathbb P}^2$    
coincides with the coefficient of ${\cal E}$ in $T^n({\cal E})$ 
as an action on Pic($X$).
Because the coefficients of
\begin{eqnarray}\label{ce} 
&&T^{sl}({\cal E})\\
&=&\sum_i d_iD_i + \sum_j (r_j+s z_j e)\alpha_j+
+ \left(sl\sum_j k_j r_j + k + \frac{1}{2} s(s-1)le\sum_jk_jz_j\right)K_X 
   +E_9, \nonumber 
\end{eqnarray}
where $n=sl$,
increase at most with the order $s^2$,
the coefficient of ${\cal E}$  also increases at most $O(n^2)$.
\ep\\

{\noindent{\bf Acknowledgment.}}
The author would like to thank H. Sakai, J. Satsuma, T. Tokihiro, K.Okamoto, 
R. Willox, A. Nobe, T. Tsuda and M. Eguchi 
for discussions and advice. \\

\appendix

\section{The space of initial values constructed by blowing up ${\mathbb P}^2$}

We consider the construction of the space of initial values for 
the HV eq. $\varphi:{\mathbb P}^2\to {\mathbb P}^2$
$$\varphi:(X,Y,Z)\mapsto (Y^3,-XY^2+Y^3+aZ^3,Y^2Z).$$
This mapping is reduced from (\ref{hv}) by the change of variables 
$x=X/Z,y=Y/Z$.

The space of initial values $X'$ becomes isomorphic to $X$ where $X$ is 
the space of initial values in the case of 
${\mathbb P}^1\times {\mathbb P}^1$. Denoting 
a class of the curve $aX+bY+cZ=0$ by ${\cal E}$, where $a,b,c\in{\mathbb C}$
($a$ bears no relation with $a$ in $\varphi$), we have a correspondence
between the bases of their Picard groups as follows
$${\rm Pic}(X')= {\bf Z}{\cal E}+{\bf Z}E_p'+{\bf Z}E_q'+{\bf Z}E_1'+\cdots+
{\bf Z}E_8'+{\bf Z}E_{10}'+{\bf Z}E_{11}'+{\bf Z}E_{14}'$$
\begin{eqnarray*}
{\cal E}=H_0+H_1-E_9, 
E_p'=H_0-E_9, E_q'=H_1-E_9,\\
E_i'=E_i \quad \mbox{for } i=1,2,3,4,5,6,7,8,10,11,12,13,14.
\end{eqnarray*}
The intersection form is
$${\cal E}\cdot{\cal E}=1,~~ E_i'\cdot E_j'=-\delta_{i,j},
~~{\cal E}\cdot E_i'=0  $$
for all $i,j\in\{p,q,1,2,\cdots,\check{9},\cdots,14\}$.\\
The irreducible components of the anti-canonical divisor are
\begin{eqnarray*}
&E_1'-E_2',~E_2'-E_3',~E_3'-E_4',~E_5'-E_6',~
E_6'-E_7',~E_7'-E_8',\\
&{\cal E}-E_p'-E_q'-E_{10}',~E_{11}'-E_{12}',
~E_{12}'-E_{13}',~E_{13}'-E_{14}'\\
&E_p'-E_1'-E_2',~E_q'-E_5'-E_6',
~E_{10}'-E_{11}'-E_{12}'
\end{eqnarray*}
and the root basis is
\begin{eqnarray*}
\alpha_1&=&2{\cal E}-2E_p'-E_1'-E_2'-E_3'-E_4',\\
\alpha_2&=&2{\cal E}-2E_q'-E_5'-E_6'-E_7'-E_8',\\
\alpha_3&=&2{\cal E}-2E_{10}'-E_{11}'-E_{12}'-E_{13}'-E_{14}'.
\end{eqnarray*}

%%%%%%%%%%%%%%%%%%%%%%%%%%%

\end{document}